\documentclass[final,5p,sort&compress]{elsarticle}

\usepackage{amssymb}
\usepackage{amsmath}

\usepackage{multirow}
\usepackage{threeparttable}
\usepackage{adjustbox}
\usepackage{siunitx}

\usepackage{lineno}

\usepackage[colorlinks]{hyperref}
\usepackage{todonotes}

\journal{NIMB}

\newcommand{\nunrt}{\nu_{d}}
\newcommand{\nuarc}{\nu_{d,\text{arc}}}

\begin{document}

\begin{frontmatter}



\title{On the use of SRIM for calculating arc-dpa exposure\\
{\small 
\textit{Published in Nucl. Instrum. Methods Phys. Res., Sect. B,} 
\href{https://doi.org/10.1016/j.nimb.2023.165145}{doi:10.1016/j.nimb.2023.165145} 
} 
}


\author[1,2]{E. Mitsi\corref{cor1}}
\ead{elmitsi@ipta.demokritos.gr}
\author[2]{K. Koutsomitis}
\author[2]{G. Apostolopoulos}

\address[1]{Department of Physics, National Technical University of Athens, Zografou Campus, EL-15780 Athens, Greece}
\address[2]{Institute of Nuclear and Radiological Science and Technology, Energy, and Safety,\\  N.C.S.R. ``Demokritos'', EL-15310 Agia Paraskevi, Greece}
\cortext[cor1]{Corresponding author}

\begin{abstract}
We propose two methods for evaluating athermal recombination corrected (arc) displacement damage parameters in ion irradiations employing the computer code SRIM (Stopping and Range of Ions in Matter). The first method consists of post-processing the detailed SRIM output for all simulated damage events and re-calculating according to the arc damage model. In the second method, an approximate empirical formula is devised which gives the average displacements in the arc damage model as a function of the corresponding quantity according to the standard Norgett-Robinson-Torrens model, which is readily obtained from SRIM. 

\end{abstract}




\end{frontmatter}



\section{Introduction}

In studies of radiation effects in materials it is generally desirable to have a standardized parameter to quantify radiation damage exposure, that would provide a
common basis for comparison of data obtained under different irradiation conditions in terms of impinging particle type and energy. 
Currently, the internationally accepted standard parameter for this purpose is the number of displacements per atom (dpa) 
calculated according to the Norgett-Robinson-Torrens (NRT) model \cite{Norgett-1975-ID480}. 
Recently, a modification to the NRT model has been proposed, the athermal recombination corrected dpa (arc-dpa) \cite{Nordlund-2015-ID597, Nordlund-2018-Improvingatomicdis, Nordlund-2018-ID1137}. It addresses a well known issue of NRT, namely, the overestimation of the number of stable defects generated by high energy displacement cascades. The arc-dpa model is based on evidence from experimental studies and computer simulations, which indicates that significant defect recombination takes place during the cascade cool-down phase leading to reduced numbers of remaining stable defects.
In its present formulation, the model can be applied to a limited set of monoatomic metallic target materials, for which the required material-specific model parameters have been obtained. It is expected that in the future, as new experimental and simulation data becomes available, the arc-dpa model will be applicable to a wider range of metallic materials including both dilute and concentrated alloys \cite{Nordlund-2018-Improvingatomicdis}.

There are several software tools available for estimating radiation damage exposure. In the case of ion irradiation, one of the most widely used such tools is the Monte Carlo code SRIM (Stopping and Range of Ions in Matter) \cite{Ziegler-2010-ID479}. Its popularity is based on the fact that it employs accurate ion stopping powers and provides a user-friendly interface. Several authors \cite{Stoller-2013-ID110,Li-2015-IM3DAparallelMon,Weber-2019-ID1179,Crocombette-2019-Quickcalculationof,Stoller-2019-ID1182,Agarwal-2021-OntheuseofSRIMf}
have discussed the application of SRIM for accurate damage calculations. The program offers different options for the simulation of damage events. Many authors have noted that the option ``Quick calculation of damage'' (Q-C) may be preferable in order to obtain results comparable to the NRT model \cite{Stoller-2013-ID110,Weber-2019-ID1179}. However, the Q-C mode implements certain approximations and thus may be less accurate, especially for multi-elemental targets \cite{Weber-2019-ID1179,Crocombette-2019-Quickcalculationof}. On the other hand, the more detailed ``Full damage cascades'' (F-C) option tends to significantly overestimate damage production compared to the Q-C mode \cite{Stoller-2013-ID110,Agarwal-2021-OntheuseofSRIMf}. \citet{Agarwal-2021-OntheuseofSRIMf} suggested employing SRIM in F-C mode to obtain the average of the damage energy, $T_d$, i.e., the ion energy deposited to target displacements. The average $T_d$ is then inserted in the NRT formula to calculate the average damage produced. This is most accurate at high damage energies, where the NRT damage function is linear with respect to $T_d$.

Regarding the arc-dpa model, there is currently no standardized way to compute damage exposure in ion irradiations as the model has not yet been implemented in any of the widely used software tools. 
Since the arc-dpa damage function is strongly non-linear, knowledge of just the average value of $T_d$ is not enough to correctly estimate the damage, as done, e.g., for NRT-dpa with the SRIM damage energy method of \citet{Agarwal-2021-OntheuseofSRIMf}.  
In their original publication introducing the new damage model, \citet{Nordlund-2015-ID597} already discussed the application of SRIM for indirect estimation of arc-dpa damage parameters. 
They presented a method of calculation in two steps. 
First, a series of SRIM simulations were performed to evaluate $T_d$ as a function of the initial PKA recoil energy, $E_R$, for a given target material. In \cite{Nordlund-2015-ID597}, this was done for Fe and an interpolating function was devised to obtain $T_d$ continuously as a function of recoil energy up to 300~keV. 
In the second part of the calculation, the information obtained on $T_d$ is used for post-processing the SRIM output file "COLLISON.txt" to finally obtain the arc-dpa values.

In this paper, we propose two alternative methods to calculate arc-dpa exposure using SRIM. 
They are based on the ``Quick calculation of damage'' (Q-C) option, which provides a NRT-compatible damage estimate as a starting point. Since the arc-dpa model refers currently only to monoatomic metals, the known limitation of the Q-C mode regarding multi-elemental targets is not relevant.   
The first of the proposed methods utilizes the SRIM output file COLLISON.txt as previously discussed in \cite{Nordlund-2015-ID597}. However, instead of separately computing $T_d$ by interpolation, we use the damage energy values that are internally calculated by SRIM with the Lind­hard-Scharff-Schiøtt (LSS) approximation \cite{Lindhard-1963-RANGECONCEPTSANDH}. Thus, the damage energy interpolation for different target materials is not required. The second method is based on an approximate formula that we propose, which can be employed for direct estimation of arc-dpa exposure based on the corresponding NRT-dpa value. Thus, the cumbersome handling of the COLLISON.txt file is avoided. The two methods are tested on all targets for which arc-dpa model parameters are available and for a range of projectile ions. 

\section{Radiation Damage Models}
The NRT model gives the number of stable displacements, $\nunrt$, produced by a PKA recoil with damage energy $T_d$ as:
\begin{equation}\label{eq:NRT}
\nunrt(T_d)=
	\begin{cases}
		  0             & \text{for } T_d \le E_d\\
		  1 			& \text{for } E_d < T_d \le L\\
T_d/L & \text{for } T_d > L
	\end{cases}	
\end{equation} \\
where $E_d$ is the displacement threshold energy, i.e., the minimum energy required to displace an atom from its lattice position. $L=2E_d/0.8$ denotes the cascade multiplication threshold above which more than one stable displacements are generated by the PKA.

In the arc-dpa model, the 3$^{rd}$ branch of \eqref{eq:NRT} is multiplied by an energy dependent efficiency factor, $\xi \leq 1$. The model definition is summarized in the following two relations:
\begin{equation}\label{eq:arc}
	\nuarc(T_d)=
	\begin{cases}
		0             & \text{for } T_d \le E_d,\\
		1 			& \text{for } E_d < T_d \le L,\\
	\xi(T_d/L)\cdot T_d/L & \text{for } T_d>L,
	\end{cases}	
\end{equation}
\begin{equation}\label{eq:xi}
	\xi(x) = (1-c)\, x^b + c, \quad \text{for } x \geq 1.
\end{equation}
The parameters \textit{b} and \textit{c} are material constants that have been determined for a number of target materials by \citet{Nordlund-2018-Improvingatomicdis}. Their values are given in table \ref{tab:parameters}.

We note that for damage energies above the displacement threshold, $T_d>E_d$, $\nuarc(T_d)$ can be compactly written as
\begin{equation}\label{eq:arc2}
\nuarc(T_d) = \nunrt(T_d)\cdot \xi\left[ \nunrt(T_d) \right].
\end{equation}
This definition will be utilized in the following paragraphs.

\section{SRIM simulation conditions and data handling} \label{SRIMcondition}

\begin{table} [tb!]
	\centering
	\begin{tabular}{cccccc}
		\hline\hline
		Ion & H & He & Al & Fe & Au\\
		\hline
		$E_0$ (MeV) & 1 & 1 & 3 & 5 & 10\\ 
		\hline\hline	
	\end{tabular}
	\caption{Projectile ions and corresponding incident energies $E_0$.}
	\label{tab:ions}
\end{table}

\begin{table} [tb!]
	\centering
	\begin{tabular}{ccccccc}
		\hline\hline
		Target & Fe & Ni & Cu & Pd & W & Pt\\
		\hline
		$E_d$ (eV) \cite{Nordlund-2015-ID597} & 40 & 40 & 29 & 41 & 90 & 44 \\
		$b$ \cite{Nordlund-2018-Improvingatomicdis} & -0.568 & -1.01 & -0.68 & -0.88 & -0.56 & -1.12\\ 
		$c$ \cite{Nordlund-2018-Improvingatomicdis} & 0.286 & 0.23 & 0.16 & 0.15 & 0.12 & 0.11\\
		\hline\hline	
	\end{tabular}
	\caption{Displacement threshold, $E_d$, and arc-dpa model parameters, $(b, c)$, of simulated targets.}
	\label{tab:parameters}
\end{table} 

\renewcommand{\arraystretch}{1.75}

\begin{table*} [tb!]
	\centering

	\begin{adjustbox}{max width=\textwidth}
	\begin{threeparttable}[t]
		
		\begin{tabular}{lccc}
			\hline\hline
			Quantity & Symbol 
			& \parbox[c][1cm]{3cm}{
				Method 1 (M1)\\COLLISON.txt
			}
			& \parbox[c]{3cm}{ \centering
				Method 2 (M2)\\VACANCY.txt
			} 
			\\
			
			\hline
			
			PKAs per ion & $N_{\text{PKA}}$ 
			& $N_{\text{rows}} / N_{\text{ions}}$ 
			& $\sum_k{ \left[ \nu_i \right]_k\, \Delta x}$ \tnote{\dag} 
			\\
			
			\hline
			
			& &\multicolumn{2}{c}{NRT-dpa model} \\
			
			Displacements per ion & $N_{d}$ 
			& $N_{\text{ions}}^{-1}\,\sum_k{\left[ \nu_d\right]_k}$ \tnote{\ddag}
			& $\sum_k{\left[ \nu_i + \nu_r \right]_k\, \Delta x} $ \tnote{\dag} 
			\\
								
			Mean displacements per PKA & $ \langle \nu_d \rangle $ 
			& \multicolumn{2}{c}{$ N_d / N_{\text{PKA}}$}\\
			
			\hline
			
			& &\multicolumn{2}{c}{arc-dpa model} \\
			
			Displacements per ion & $N_{d,arc}$ 
			& \multicolumn{2}{c}{$\langle \nu_{d,arc} \rangle \cdot N_{\text{PKA}} $}\\
			
			Mean displacements per PKA & $ \langle \nu_{d,arc} \rangle $ 
			& $N_{\text{rows}}^{-1}\,\sum_k{
				\left[ \nu_d\right]_k \xi\left( \left[ \nu_d\right]_k \right)
				}$ \tnote{\ddag}
			& \parbox[c]{3cm}{
			\begin{center}
				eq. \eqref{eq:ndarc_approx} with $\langle \nu_d \rangle $\\as above
			\end{center}}
			\\

			\hline\hline	
		\end{tabular}

		\begin{tablenotes}
			\item [\dag]  $\left[ \nu_i \right]_k$ and $\left[ \nu_r \right]_k$ are the ``vacancies by ions'' 
			              and ``vacancies by recoils'', respectively, in the $k$-th target depth bin, with  
						  $\Delta x$ denoting the bin width.
			\item [\ddag]  $\left[ \nu_d \right]_k$ denotes the number of vacancies 
			               estimated by SRIM for the $k$-th PKA event. The sum is over all events.
			
		\end{tablenotes}

	\end{threeparttable}
	\end{adjustbox}
	\caption{Calculation of damage parameters from SRIM output files}
	\label{tab:quantities}
\end{table*} 
\renewcommand{\arraystretch}{1.0}

All simulations were performed utilizing SRIM-2013 and employing the option "Ion distribution and Quick calculation of damage" (Q-C). 
A range of projectile ions were employed, with atomic numbers varying from $Z=1$ (H) to 79 (Au)
and energies ranging from $E_0=1$ to 10 MeV, similarly to the work of \citet{Agarwal-2021-OntheuseofSRIMf}. The ions and corresponding energies are listed in
Table \ref{tab:ions}. Table \ref{tab:parameters} shows all the targets that we tested, which are essentially all materials whose arc-dpa parameters were estimated in \cite{Nordlund-2018-Improvingatomicdis}. Target
thickness was chosen appropriately in order to ensure that
the impinging ions stop within the examined
region. 
The target displacement energies, $E_d$, are based on internationally recommended standard values and are also given in Table \ref{tab:parameters}.
In the case of Fe self-ion irradiation, an extra simulation with $E_0=78.7$~keV was also performed in order to directly compare with results from \cite{Nordlund-2015-ID597}. For each ion/target combination 10,000 ion histories were run.

Damage parameters are extracted from the SRIM output files, either VACANCY.txt or COLLISON.txt. Table \ref{tab:quantities} lists all quantities of interest and the way they are calculated depending on the damage model and the output file used. 

The number of PKAs per ion, $N_{\text{PKA}}$, is obtained by integrating the 2nd data column of VACANCY.txt (``vacancies by ions'',  $\nu_i$) or by dividing the number of data rows, $N_{\text{rows}}$, in COLLISON.txt by the number of simulated ions, $N_{\text{ions}}$. $N_{\text{PKA}}$ is independent of the damage model. 

The NRT displacements per ion, $N_d$, is obtained as follows. In the case of VACANCY.txt, $N_d$ is found by summing the $2^{nd}$ and $3^{rd}$ column of the data table, i.e., ``vacancies by ions'', $\nu_i$, and "vacancies by recoils", $\nu_r$, respectively, as suggested by previous authors \cite{Stoller-2013-ID110,Nordlund-2015-ID597}. 
Regarding the COLLISON.txt file, $N_d$ is calculated by adding up the "Target vacancies",  $\nu_d$, of all PKAs and dividing by $N_{\text{ions}}$. Finally, the average displacements per PKA, $\langle \nu_d \rangle$, is equal to $N_d / N_{\text{PKA}}$. 

The calculation of arc-dpa damage parameters is described in the next section.

All evaluations and the parsing of SRIM output files were performed in the OCTAVE computing environment \cite{Eaton2022}. The open source python code PYSRIM \cite{pysrim} was employed to automate the SRIM calculations. All relevant data and code are available in \cite{mitsi_e_2023_8116031}.

\section{Methods}
In this section, we present the two different methods to obtain arc-dpa damage parameters from SRIM output.

\subsection{Method 1 (M1)}
This method utilizes the COLLISON.txt output file. In SRIM Q-C mode, this file lists all simulated PKA scattering events and reports, among other data, the number of displacements, $\nu_d$, generated per event. These $\nu_d$ values, labelled "Target vacancies", are calculated according to the NRT model, eq. \eqref{eq:NRT}, with the damage energy, $T_d$, obtained from the approximate LSS theory \cite{Ziegler-2008-ID880}. For $\nu_d > 1$, we can easily recover the LSS damage energy by multiplying $\nunrt$ with the cascade multiplication factor, $L$ (cf. eq. \eqref{eq:NRT}). Then, the obtained $T_d$ can be used in eq. \eqref{eq:arc} to evaluate the displacements according to the arc-dpa model. Equivalently, $\nuarc$ can be obtained by plugging $\nunrt$ directly into the alternative arc-dpa definition, eq. \eqref{eq:arc2}. The steps to calculate the arc-dpa damage parameters are as follows: 
\begin{enumerate}
	\item Run SRIM with the "Quick calculation of damage" (Q-C) option.
	\item Parse the COLLISON.txt output file to obtain the NRT displacements per PKA event, $\nunrt$.
	\item Calculate the corresponding $\nuarc$ per PKA from eq. \eqref{eq:arc2}, $\nuarc = \nunrt\cdot \xi(\nunrt)$.
	\item Take the average of the $\nuarc$ values to obtain the mean displacements per PKA according to the arc-dpa model, $\langle\nuarc\rangle$ (cf. Table \ref{tab:quantities}).
	\item Multiply by the number of PKAs per ion, $N_{PKA}$, to obtain the number of displacements per ion, $N_{d,arc} = \langle\nuarc\rangle \cdot N_{\text{PKA}}$
\end{enumerate}

\subsection*{Method 2 (M2)}

\begin{figure}[!tb]
	\centering
	\includegraphics[width=88mm]{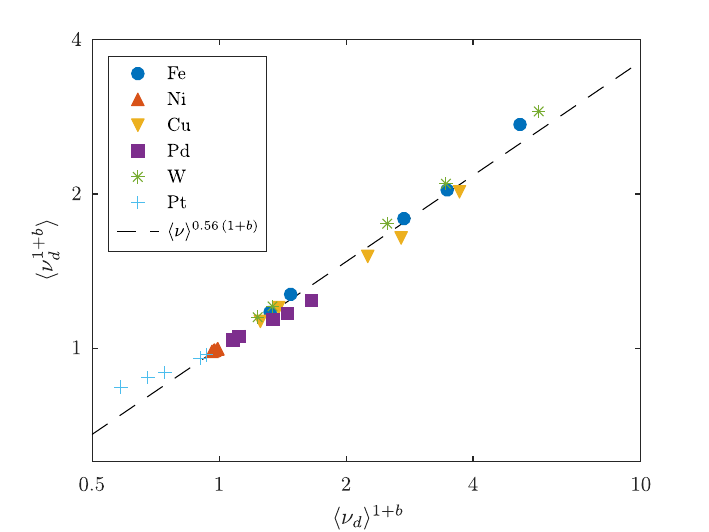}
	\caption{
	$ \langle\nunrt^{1+b}\rangle $ as a function of 
	$ \langle\nunrt\rangle^{1+b} $, where $\nu_d$ denotes the NRT displacements and $b$ is the arc-dpa model parameter of the corresponding target material. Both quantities were obtained by post-processing the output of SRIM simulations and averaging over all PKA events. Results for the different target materials are depicted with different symbol and color. The dashed line corresponds to the approximate relation \eqref{eq:nuapprox}.
	}
	\label{fig1}
\end{figure}

The objective of M2 is to provide a quick estimate of the arc-dpa damage parameters, without having to resort to the cumbersome processing of COLLISON.txt. For this, we note that from eq. \eqref{eq:arc2} the average arc-dpa can be written:
\begin{equation}\label{eq:arc3}
\langle\nuarc\rangle = (1-c)\langle\nunrt^{1+b}\rangle + c \cdot \langle\nunrt\rangle.
\end{equation}
Thus, to obtain $\langle\nuarc\rangle$ the value of $\langle\nunrt^{1+b}\rangle$ is needed. We performed an approximate calculation of this quantity, employing a power-law cross-section for the ion-target atom interaction and ignoring the effect of ionization losses, i.e., setting $T_d \approx T$. As shown in \ref{apdx1}, the following approximation
\begin{equation}\label{eq:nuapprox}
\langle\nunrt^{1+b}\rangle \approx \langle\nunrt\rangle^{\lambda (1+b)},
\end{equation}
where $ \lambda\approx 0.56 $, gives adequate results for a wide range of incident ion energies and ion-target combinations. 
This can be seen in fig. \ref{fig1}, where
$ \langle\nunrt^{1+b}\rangle $ 
is plotted as a function of 
$ \langle\nunrt\rangle^{1+b} $ for all the ion/target combinations simulated in the current work. The data shown in the figure have been obtained by taking the $\nunrt$ values per PKA event listed in COLLISON.txt and evaluating the required averages. As seen from the figure, the data from all simulated targets lie within $\pm 10\%$ of the approximate eq. \eqref{eq:nuapprox}, which is depicted by the dashed line.

Utilizing the above approximation, the arc-dpa damage parameters can be obtained by the following prescription:
\begin{enumerate}
	\item Run SRIM with the "Quick calculation of damage" (Q-C) option.
	\item Calculate the NRT-$\langle \nunrt \rangle$ from VACANCY.txt as described in Table \ref{tab:quantities}.
	\item Obtain $\langle\nuarc\rangle$ from eq. \eqref{eq:arc3}, substituting the approximate relation \eqref{eq:nuapprox}:
	\begin{equation}\label{eq:ndarc_approx}
	\langle\nuarc\rangle \approx (1-c)\langle\nunrt\rangle^{0.56(1+b)} + c \cdot \langle\nunrt\rangle
	\end{equation}
	\item The number of displacements per ion is $\langle\nu_{d,arc}\rangle \cdot N_{\text{PKA}}$
\end{enumerate}

\section{Results and discussion}

\begin{table*} [tb!]
	\centering
	\begin{threeparttable}[t]
	\begin{adjustbox}{max width=\textwidth}
	\begin{tabular}{lcccScS}
		\hline\hline
		{} & $E_0$ & $N_{\text{PKA}}$ & {$N_{d}$} & {$\langle\nunrt\rangle$} 
		& {$N_{d, arc}$}  & {$\langle\nuarc\rangle$} \\
		\hline
		\citet{Nordlund-2015-ID597} & \multirow{3}{*}{78.7 keV} & \multirow{3}{*}{44.1} & 
		539 & 12.2\tnote{\dag} & 217 & 4.93\tnote{\dag} \\
		This study - Method 1 & {} & {} & 
		530 & 12.0 & 217 & 4.92 \\ 
		This study - Method 2 & {} & {} & 
		530 & 12.0 & 209 & 4.74 \\
		\hline
		Method of \citet{Nordlund-2015-ID597} & \multirow{3}{*}{5 MeV} & \multirow{3}{*}{442} & 
		8800 & 20.0 & 3150 & 7.14 \\
		This study - Method 1 & {} & {} & 
		7870 & 17.9 & 2900 & 6.56 \\ 
		This study - Method 2 & {} & {} & 
		7870 & 17.9 & 2890 & 6.54 \\
		\hline\hline	
	\end{tabular}
	\end{adjustbox}

	\begin{tablenotes}
		\item [\dag] $\langle\nunrt\rangle$ and $\langle\nuarc\rangle$ are calculated by dividing $N_d$ and $N_{d,arc}$ from \cite{Nordlund-2015-ID597}, respectively, by $N_{\text{PKA}}$ as obtained in the present study.
		 
	\end{tablenotes}
	\end{threeparttable}

	\caption{Damage parameters obtained by different methods for the irradiation of an Fe target with Fe ions of energy $E_0$.}
	\label{tab:comparison}
\end{table*}

Damage parameters obtained by SRIM according to method M1 are compared to the results of \citet{Nordlund-2015-ID597}.
First, we repeated the simulation of 78.7 keV Fe ions incident on an Fe target that was reported in \cite{Nordlund-2015-ID597}. The results of both methods are given in Table \ref{tab:comparison}.
As seen from the table, there is a small 2\% difference in the NRT parameters, $N_d$ and $\langle\nunrt\rangle$, between our M1 and the results of \cite{Nordlund-2015-ID597}. The corresponding arc-dpa parameters almost coincide. 
As the damage energies occurring in this example are relatively low, we simulated self-ion Fe irradiation with a much higher projectile energy, $E_0=5$~MeV, and evaluated the results with both our proposed method M1 and the one described in \citet{Nordlund-2015-ID597}.  
In the latter case, we used the data from their fig.~1.2 to extend the interpolation of $T_d$ to target recoil energies up to 5~MeV. 
The resulting damage parameters, also listed in Table \ref{tab:comparison}, show that there is a 10\% difference between the NRT parameters obtained by our method M1 and the evaluation according to \cite{Nordlund-2015-ID597}. 
The corresponding arc-dpa parameters exhibit a similar but slightly lower discrepancy of about 8\%.
We attribute the differences in damage parameters to the distinct way the damage energy is evaluated in the two methods. In the present work we employ the Q-C mode, while \citet{Nordlund-2015-ID597} used SRIM's "Detailed Calculation with Full Damage Cascades" (F-C) option. In the latter case, SRIM utilizes detailed stopping power calculations for all secondary recoils in the PKA cascade, thus, the value of $T_d$ is potentially more accurate than in the Q-C mode where the LSS approximation is employed. \citet{Agarwal-2021-OntheuseofSRIMf} have made a detailed comparison of SRIM damage calculations in Q-C and F-C modes. They found differences up to $\pm 25\%$ in the damage energy predicted by the two modes. The discrepancies we observe here between M1 and the method of \cite{Nordlund-2015-ID597} are of comparable magnitude. The smaller discrepancy observed in the arc-dpa parameters is due to the fact that the arc-dpa damage function lowers the significance of high energy damage events, where the errors due to the LSS approximation are more pronounced.

\begin{table*} [tb!]
	\centering
	\begin{adjustbox}{max width=\textwidth}
	\begin{tabular}{lSSS@{\hspace{1cm}}SSS@{\hspace{1cm}}SSS}
		\hline\hline
		Projectile &
		$\langle\nunrt\rangle$ & 
		\multicolumn{2}{c@{\hspace{1cm}}}{$\langle\nuarc\rangle$} & 
			$\langle\nunrt\rangle$ & 
		\multicolumn{2}{c@{\hspace{1cm}}}{$\langle\nuarc\rangle$} & 
		$\langle\nunrt\rangle$ & 
		\multicolumn{2}{c}{$\langle\nuarc\rangle$} \\
		
		{} & {}
		& {M1} & {M2} &
		& {M1} & {M2} &
		& {M1} & {M2} \\
		\hline
		& \multicolumn{3}{c@{\hspace{1cm}}}{Fe target}
		& \multicolumn{3}{c@{\hspace{1cm}}}{Ni target}
		& \multicolumn{3}{c}{Cu target} \\		
	
		1 MeV H & 1.91 & 1.38 & 1.38 & 1.92 & 1.21 & 1.21 & 2.02 & 1.27 & 1.28 \\
		1 MeV He & 2.47 & 1.62 & 1.60 & 2.54 & 1.35 & 1.35 & 2.76 & 1.45 & 1.45 \\
		3 MeV Al  & 10.4 & 4.24 & 4.22 & 10.6 & 3.21 & 3.21 & 12.6 & 3.29 & 3.35 \\
		5 MeV Fe & 17.9 & 6.56 & 6.54 & 18.6 & 5.04 & 5.04 & 22.3 & 4.95 & 5.04 \\
		10 MeV Au  & 44.8 & 14.8 & 14.6 & 47.7 & 11.7 & 11.7 & 60.4 & 11.4 & 11.4 \\
		
		\hline
		& \multicolumn{3}{c@{\hspace{1cm}}}{Pd target}
		& \multicolumn{3}{c@{\hspace{1cm}}}{W target}
		& \multicolumn{3}{c}{Pt target} \\
			
		1 MeV H & 1.89 & 1.17 & 1.17 & 1.61 & 1.20 & 1.18 & 1.78 & 1.06 & 1.05  \\
		1 MeV He & 2.47 & 1.27 & 1.27 & 1.94 & 1.29 & 1.27 & 2.37 & 1.11 & 1.10  \\
		3 MeV Al & 11.6 & 2.70 & 2.73 & 8.06 & 2.51 & 2.44 & 11.7 & 2.08 & 2.04  \\
		5 MeV Fe & 22.7 & 4.39 & 4.45 & 16.7 & 3.85 & 3.76 & 25.6 & 3.59 & 3.53  \\
		10 MeV Au & 67.0 & 11.1 & 11.2 & 52.8 & 8.88 & 8.67 & 87.4 & 10.4 & 10.3  \\

		\hline\hline
	\end{tabular}
	\end{adjustbox}			
	\caption{Mean displacements per primary knock-on atom for both NRT- and arc-dpa models, $\langle\nunrt\rangle$ and $\langle\nuarc\rangle$, respectively, as obtained by SRIM. For $\langle\nuarc\rangle$ the results of both calculation methods M1 and M2 are given. }
	\label{tab:all_data}
\end{table*} 

\begin{figure}[!tb]
	\centering
	\includegraphics[width=88mm]{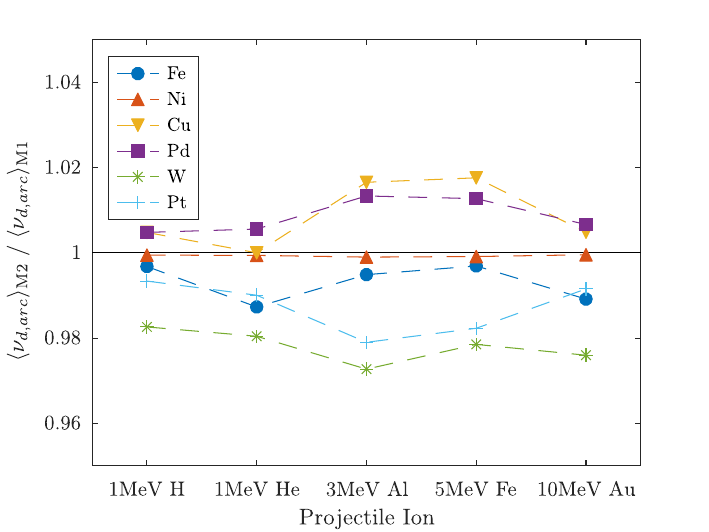}
	\caption{The ratio of the average displacements per PKA according to the arc-dpa model, $\langle\nuarc\rangle$, obtained by methods M2 and M1. Results for the different target materials are depicted with different symbol and color.}
	\label{fig2}
\end{figure}

Table \ref{tab:comparison} shows also the results of the approximate method M2 for the two Fe self-irradiation simulations. The NRT damage parameters of methods M1 and M2, obtained from the files COLLISON.txt and VACANCY.txt, respectively, are identical as expected. The corresponding arc-dpa parameters exhibit a discrepancy of 4\% and 0.3\% in the low and high energy simulation, respectively. Table \ref{tab:all_data} lists the values of NRT mean displacements per PKA, $\langle \nunrt \rangle$, and the corresponding arc-dpa value, $\langle \nuarc \rangle$, as calculated by both methods M1 and M2 for all test cases that we simulated in the current work. 
Comparing the two models, it is observed that the arc-dpa predicts significantly fewer displacements per PKA at high projectile energies compared to NRT, while at low projectile energies the difference is not so large. This is exactly the behavior anticipated for the arc-dpa model \cite{Nordlund-2015-ID597,Nordlund-2018-Improvingatomicdis,Nordlund-2018-ID1137}. The values of $\langle \nuarc \rangle$ calculated by methods M1 and M2 are in all cases very similar.
Fig. \ref{fig2} depicts the ratio of $\langle\nuarc\rangle$ obtained by M2 and M1, respectively. As seen from the figure, the approximate method M2 deviates by at most 3\% from the results of M1.
Thus, the method M2 can be employed for an approximate evaluation of arc-dpa damage, introducing an error of not more than a few percent compared to the more detailed method M1. 

\subsection{Depth-dependent calculations}
In many applications the depth-dependent damage profile is also of interest. Both methods M1 and M2 can be employed for obtaining the arc-dpa damage profile from SRIM output.
This is most straightforward in the case of M2, were the calculations shown in Table \ref{tab:quantities} can be applied line-by-line to the data of VACANCY.txt and thus obtain $\langle \nuarc \rangle$ and $N_{d,arc}$ as a function of depth.
On the other hand, for method M1 extra processing is required to select from COLLISON.txt those PKA events which occur within a certain target depth bin and then perform the calculations of Table \ref{tab:quantities} to obtain the arc-dpa parameters in this particular bin. By iterating this procedure over all depth bins we finally obtain the damage profile.

\begin{figure}[!tb]
	\centering
	\includegraphics[width=88mm]{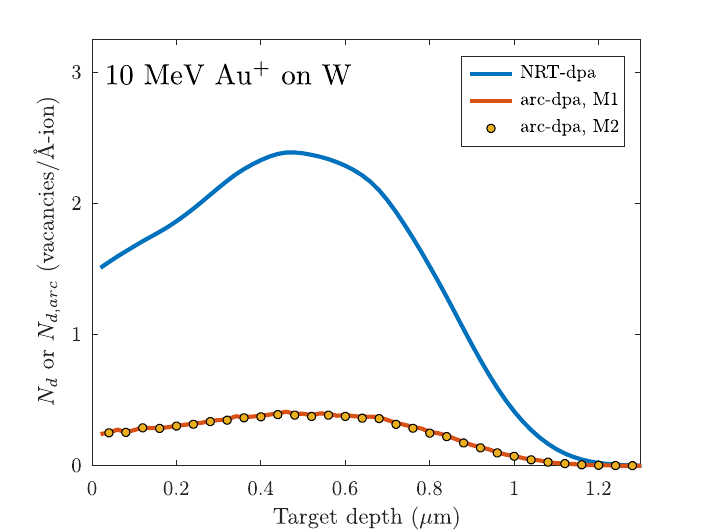}
	\caption{Damage profile for the irradiation of W by 10~MeV Au ions according to the NRT- and arc-dpa model as obtained by SRIM. The arc-dpa profile is calculated by both methods M1 (continuous curve) and M2 (circles).}
	\label{AuonW}
\end{figure}

Indicative results of depth-dependent application of methods M1 and M2 are depicted in Fig. \ref{AuonW}. The figure shows the damage profiles predicted by SRIM for a 10~MeV Au irradiation of W in the standard NRT-dpa model and the arc-dpa model as obtained by methods M1 and M2. As seen in the figure, there is a peak in vacancy production at about 0.5~$\mu$m in both NRT- and arc-dpa data. The arc-dpa profiles obtained by M1 and M2 are almost identical.

\section{Conclusions}

In this work, we present two methods for evaluating arc-dpa damage parameters in ion irradiations employing the SRIM simulation code. The methods are based on the ``Quick calculation of damage'' (Q-C) option to obtain an initial estimate of displacement damage compatible with the NRT standard. 

The first method employs SRIM's COLLISON.txt output file, which lists the NRT displacements, $\nunrt$, produced in each simulated primary knock-on atom (PKA) recoil event. The $\nunrt$ values are converted to the corresponding arc-dpa model prediction, $\nuarc$, by means of eq. \eqref{eq:arc2} and then averaged to obtain the total damage parameters. This procedure is similar to the one proposed by \citet{Nordlund-2015-ID597} only in our case the damage energy, $T_d$, is essentially obtained by the LSS approximation employed in SRIM's Q-C mode, whereas in \cite{Nordlund-2015-ID597} the damage energy was interpolated from the results of separate detailed SRIM simulations. Thus, our method gains in simplicity but can lead to errors due to the approximation in the damage energy calculation. According to \citet{Agarwal-2021-OntheuseofSRIMf} the discrepancy in $T_d$ obtained by Q-C and F-C modes, respectively, could be up to $\sim 25\%$. It is expected that this would be also the upper limit for the discrepancy in arc-dpa damage. 

In the second method, we devise an approximate relation, which gives $\langle\nuarc\rangle$ directly as a function of $\langle\nunrt\rangle$. Thus, the cumbersome processing of the COLLISON.txt file is not needed since $\langle\nunrt\rangle$ can be easily obtained from VACANCY.txt. 
We found that the arc-dpa parameters obtained by this approximate method differ by not more than a few percent from those calculated by the first method. 

Both methods can also be employed for depth-dependent arc-dpa damage calculations. 

Finally, it is noted that if the arc-dpa model is expanded to more complex systems in the future, as, e.g., concentrated alloys, the use of the ``quick damage'' option may have to be re-evaluated, as it does not handle properly multi-elemental targets.

\section*{Acknowledgement}

This work has been carried out within the framework of the EUROfusion Consortium, funded by the European Union via the Euratom Research and Training Programme (Grant Agreement No 101052200 — EUROfusion). Views and opinions expressed are however those of the author(s) only and do not necessarily reflect those of the European Union or the European Commission. Neither the European Union nor the European Commission can be held responsible for them. The funding from the Hellenic General Secretariat for Research and Innovation for the Greek National Programme of the Controlled Thermonuclear Fusion is also acknowledged.

\appendix 
\section{Approximation of $\langle \nu_d^{1+b} \rangle$}
\label{apdx1}
The general expression for the average $\langle \nu_d^n \rangle$ is given by 
\begin{equation}\label{eq:gennu}
\langle\nu_d^n\rangle = \frac {\int_{Ed}^{T_{m}} [\nu_d(T_d)]^n \,  d \sigma(E,T)}{\int_{Ed}^{T_{m}} d \sigma(E,T)},
\end{equation}
where $d\sigma(E,T)$ denotes the cross-section for scattering of an ion with initial energy $E$ producing a PKA with recoil energy $T$. $T_m$ is the maximum PKA recoil energy. Making the following assumptions:
\begin{enumerate}[(i)]
	\item A power-law cross-section, $d\sigma(E,T) \propto dT/T^{1+p}$, where $p$ ranges from 0.5 (heavy ions) to 1 (light ions) \cite{Was-2017-ID904}
	\item Ionization losses can be ignored ($T_d \approx T$)
\end{enumerate}
and performing the integrations in eq. \eqref{eq:gennu} we obtain the following analytical expression: 
\begin{equation}\label{eq:nu}
\langle\nu_d^n\rangle	= 
\frac{(L/E_d)^{p} - 1 + \frac{p}{p-n}\left[ 1 - (L/T_m)^{p-n} \right]}
{(L/E_d)^{p} - (L/T_m)^{p}} \; ,
\end{equation}
which is valid for $T_m\geq L$ and $n\neq p$. In the special case $n=p$ it becomes
\begin{equation}\label{eq:nu1}
\langle\nu_d^n\rangle	= 
\frac{(L/E_d)^{n} - 1 - n \log(L/T_m) }
{(L/E_d)^{n} - (L/T_m)^{n}} \; .
\end{equation}

\begin{figure}[h!]
\centering
\includegraphics[width=88mm]{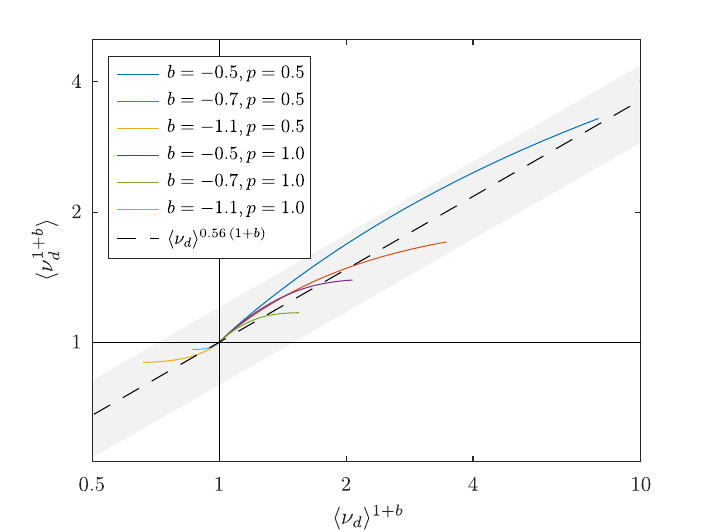}
\caption{Curves of $\langle\nu_d^{1+b}\rangle$ from eqs. \eqref{eq:nu}-\eqref{eq:nu1} as a function of $\langle\nu\rangle^{1+b}$ for different values of the parameter $b$ and the power-law cross-section exponent, $p$. The dashed line is a fit to eq. \eqref{eq:powerlaw}.}
\label{fig:nu_theory}
\end{figure}

Based on eqs. \eqref{eq:nu}-\eqref{eq:nu1} we calculate $\langle\nu_d^{1+b}\rangle$ for several representative $(b,p)$ combinations and for $T_m$ values in the range $L < T_m < 10^4 L$. This corresponds to a maximum $T_m$ of $\sim 10^6$~eV in Fe and similar values for other metals. The results are shown in fig. \ref{fig:nu_theory} as a function of $\langle\nu_d\rangle^{1+b}$ in a double-logarithmic plot. $\langle\nu_d\rangle$ is also obtained from \eqref{eq:nu}.
It is apparent from the figure that all curves follow roughly a central line. 
Fitting a power law of the form: 
\begin{equation}\label{eq:powerlaw}
\langle\nu_d^{1+b}\rangle \approx 
A \, \langle\nu_d\rangle^{\lambda\, (1+b)},
\end{equation}
to the data, with $A$ and $\lambda$ as adjustable parameters, we obtain the values $\lambda \approx 0.56$ and $A \approx 1.0$. This is denoted by the dashed line in fig. \ref{fig:nu_theory}. The deviation of the analytically calculated $\langle\nu_d^{1+b}\rangle$ from the fitted power law is within $\pm 20\%$, which corresponds to the shaded area in fig. \ref{fig:nu_theory}. 

%
%
%
%
%
%


\bibliographystyle{elsarticle-num-names} 
\bibliography{srim-arc,pysrim,zenodo}


\end{document}